\documentclass[12pt]{article}
\usepackage{a4}
\usepackage{psfig}
\usepackage{epsfig}
\usepackage{epsf}
\usepackage{rotating}
\usepackage{citesort}       
\usepackage{amssymb}
\usepackage{graphicx}

\newcommand{\vecc}[1]{\mbox{\boldmath $#1$}}
\newcommand{\vR}{\vecc{R}}           
\newcommand{\eV}{eV}

\newcommand{\text}[1]{\mbox{$\rm #1 $}}

\def\npbps#1#2#3{  { Nucl. Phys. }(Proc. Suppl.){\bf B #1} (19#2) #3}
\def\plb#1#2#3{    { Phys. Lett. }{\bf B #1} (19#2) #3}

\def\prl#1#2#3{    { Phys. Rev. Lett. }{\bf #1} (19#2) #3}

\begin{document}
\begin{flushright}
{\small
IFUM-871/FT\\
FTUAM-02-459\\
ULB-02-167
}
\end{flushright}

\vspace{0.2cm}
\begin{center}
{ \Large \bf  Neutrino mass parameters from 
Kamland, SNO and other solar evidence}\\[0.2cm]
{\large P.~Aliani$^{ab\star}$, V.~Antonelli$^{a\star}$, M.~Picariello$^{a\star}$, E.~Torrente-Lujan$^{c\star}$\\[2mm]
$^a$ {\small\sl Dip. di Fisica, Univ. di Milano},
{\small\sl and INFN Sez. Milano,  Via Celoria 16, Milano, Italy}\\
$^b$ {\small\sl Dept. Theoretical Physics, 
Univ. Libre de Bruxelles, Bruxelles, Belgium,}\\
$^c$ {\small\sl Dept. Fisica Teorica C-XI, 
Univ. Autonoma de Madrid, 28049 Madrid, Spain,}\\
}

\end{center}

\abstract{
An updated analysis of all available neutrino oscillation evidence 
 in Reactor (Kamland, first 145 days of data) and 
Solar experiments (SK day and night spectra, global rates from 
 Homestake, SAGE and GALLEX) 
 including the  $SNO  CC$ and $NC$ data 
is presented. 
In the framework of two active neutrino oscillations 
we determine the allowed regions in neutrino parameter
space,
we obtain, from the Kamland spectral shape and global signal,
the following antineutrino best solution parameters,
$ \Delta m^2_{kl}= 7.7\times 10^{-5} \eV^2, \tan^2\theta_{kl}=  0.98.$ 
The overall effect of the measured observed ratio 
$R\sim 0.6$ is that the LMA region remains the only 
one which is still favored. 

Combining Kamland and Solar data and assuming CPT invariance, 
i.e. the same mass matrix for neutrino and antineutrinos, 
we obtain
the following antineutrino best solution parameters (LMAI solution),
$ \Delta m^2= 7.1\times 10^{-5} \eV^2,\tan^2\theta=  0.47.$ 
A second solution (LMAII) appears for values 
$ \Delta m^2= 1.5\times 10^{-4} \eV^2,
\tan^2\theta=  0.48.$
We determine additionally individual neutrino mixing parameters and 
their errors from fits to marginal likelihood distributions, 
the values are compatible with previous results.
In both methods, $\chi^2$ minimization and 
marginal likelihood, the combined analysis of solar and Kamland data
concludes that  maximal mixing is not favored at the $\sim 3\sigma $ level at least. 
}

\vskip .5truecm


\vfill
{\small {}$^\star$ e-mail: paola.aliani@cern.ch, vito.antonelli@mi.infn.it, marco.picariello@mi.infn.it, emilio.torrente-lujan@cern.ch}

\newpage

\section{Introduction}
Evidence of antineutrino disappearance in a beam of antineutrinos 
in the Kamland experiment has been recently 
presented \cite{kloctober}.
The analysis of these results \cite{kloctober,klothers} 
in terms of neutrino oscillations
have   largely 
improved our knowledge of neutrino mixing on the LMA region.
The results appear to confirm in a independent way that the observed 
deficit of solar neutrinos is indeed due to neutrino oscillations. 
The ability to measure the LMA solution, the one
preferred by the solar neutrino data at present,
 ``in the lab'' puts Kamland in a pioneering situation: after these 
results there should remain little doubt of the physical reality of 
neutrino mass and oscillations.

The publication of the  SNO 
results~\cite{Ahmad:2002ka,Ahmad:2002jz}
has already 
made an important breakthrough towards the solution of the long standing
 solar neutrino 
\cite{Aliani:2002ma,Strumia:2002rv,Aliani:2002er}
problem (SNP) possible.
These results provide the strongest evidence so 
far (at least until Kamland 
improves its statistics) for flavor oscillation 
in the neutral lepton sector. 
From the combined analysis of SNO and other solar evidence, 
one obtains, 
in the framework of two active neutrino oscillations,
the following set of parameters, 
$\Delta m^2_{solar}= 4.5^{+2.7}_{-1.4}\times 10^{-5} \eV^2,
 \tan^2\theta_{solar}=  0.40^{+0.10}_{-0.08}.$ 
We will see in this work how the evidence from 
 the Kamland measurements improves or modifies these values.

The previous generation of reactor experiments 
(CHOOZ~\cite{CHOOZ}, PaloVerde~\cite{PaloVerde}), 
performed with a baseline of about 1 km.
They have attained  a sensitivity 
of $\Delta m^2<10^{-3}\ eV^2$ \cite{chooznew,CHOOZ} 
and, not finding any disappearance of the initial flux, 
they demonstrated that the atmospheric neutrino anomaly
\cite{atmospheric} is not due to muon-electron neutrino 
oscillations. The Kamland experiment is the 
successor of such experiments at a much larger scale in 
terms of baseline distance and total incident flux.  
This experiment relies upon a 1 kton liquid scintillator
 detector   located at the old, enlarged,  Kamiokande site.
 It searches for the oscillation of antineutrinos 
emitted by several nuclear power plants in Japan. 
The nearby 16  (of a total of 51) nuclear power stations deliver 
a $\overline{\nu}_e$ flux of $1.3\times 10^6 cm^{-2}s^{-1}$
for neutrino energies $E_\nu>1.8$ MeV at the detector position. 
About $85\%$ of this flux comes from  reactors forming a 
well defined baseline of 139-344 km. Thus, the flight range 
is limited  in spite of using  several reactors, because of this 
fact the sensitivity of Kamland   increases by nearly two 
orders of magnitude compared to previous reactor experiments.

The aim of this work is to study the implications of the 
recent Kamland results on the determination of the neutrino
oscillation parameters, 
 to understand  which regions
 of the parameter space still allowed by the solar 
neutrino experiments are favored by them.
The structure of this work is the following.
In section 2
 we discuss the main features of Kamland experiment 
that are relevant for our analysis:
The
salient aspects of the procedure we are adopting and 
the  results of our analysis are presented and discussed in
 section 
\ref{analysis}. Finally, 
in section \ref{conclusions} we draw our 
conclusions and discuss possible future scenarios.

\section{The computation of the expected signals}
\label{klsignal}

\subsection{The Kamland signal}
\label{Kamland}

Electron antineutrinos from nuclear reactors  with energies 
above 1.8 MeV are measured in Kamland by detecting the inverse 
$\beta$-decay reaction $\overline{\nu}_e+p\to n+e^+$. The time 
coincidence, the space correlation and the energy balance  
between the positron signal and the 2.2 MeV $\gamma$-ray
 produced by the capture of a already-thermalized  neutron on a
 free proton make it possible to identify this reaction 
unambiguously, even in the presence of a rather large background. 

The two principal ingredients in the calculation of the expected 
signal in Kamland are  the reactor  flux and
the antineutrino cross section on protons. 
A number of short baseline experiments 
(See~Ref.\cite{Murayama:2000iq} and references therein) 
have previously measured the energy spectrum of reactors at distances 
where oscillatory effects have been shown to be nonexistent. 
They have shown that the theoretical neutrino flux predictions 
are reliable within 2\% \cite{piepke}.
The effective flux of antineutrinos released by the nuclear
 plants is a rather well  understood function of 
the thermal power of the reactor and
 the amount of thermal power emitted during the 
fission of a given nucleus, which gives the total amount, and 
the  isotopic composition of the reactor fuel which gives the 
spectral shape.  
Detailed tables for these
 magnitudes can be found in Ref.~\cite{Murayama:2000iq}.
For a given isotope the energy  spectrum can be parametrized 
by an exponential expression \cite{vogel}
where the coefficients depend 
on the nature of the fissionable isotope 
(see Ref.\cite{Murayama:2000iq} for explicit values).
Along the year, between periods of refueling, the total 
effective flux changes 
with time as the fuel is expended and the isotope 
relative composition varies.
We take the average of the relative fission yields over the live 
time as given by the experiment: 
${}^{235} U=57\% $,
${}^{238} U=7.8\% $,
${}^{239} Pu=30\% $,
${}^{241} Pu=5.7\% $.

In order to obtain the expected number of events at Kamland, 
we sum the expectations for all the relevant reactor sources weighting 
each source by its power and distance to the detector
(table II in Ref.~\cite{Murayama:2000iq} ), 
 assuming the same 
spectrum originated from each reactor. 
We sum over the nearby  power reactors, we neglect 
farther Japanese and Korean reactors and even farther 
rest-of-the-world reactors which  give only a minor additional
contribution.
The average number of positrons $N_i$ which are detected per
visible energy bin $\Delta E_i$ is given by the convolution of 
different quantities: 
$\overline{P}$, the oscillation probability 
averaged  over the  distance and power of the different 
reactors.
Expressions for the 
antineutrino capture cross section  are taken from 
the literature \cite{vogel,kltorrente}. The  matrix 
element  for this cross section can 
be written in terms of the neutron half-life, 
 we have used the latest published 
value $t_{1/2}=613.9\pm 0.55$ \cite{PDG2002}.
The antineutrino flux spectrum, 
the relative reactor-reactor 
 power normalization which 
is included in the definition of 
 $\overline{P}$ and 
the energy resolution of Kamland are used in addition. 
We use in our analysis  the following
expression for the energy resolution in the prompt 
positron detection
\begin{eqnarray}
\sigma(E_e)&=&0.0062+ 0.065\surd E_e.
\label{klresolution}
\end{eqnarray}
This expression is obtained from the raw calibration 
data presented in Ref.\cite{klstony}.
Note that we prefer to use this  expression instead of the 
much less accurate one given in 
Ref.\cite{kloctober}.
Moreover, we assume a  408 ton fiducial 
mass and  standard 
nuclear plant power and fuel schedule, we take 
an averaged, time-independent, fuel composition 
equal for each detector as given above. Detection efficiency is 
taken close  100\% and independent of the energy \cite{kloctober}.

We will not add any background events as they  
 (a total of $0.95\pm 0.99$ events which 
include 
random coincidences 
from radioactive decays, $<0.01$ evts, 
and correlated background from cosmic ray muons and 
neutrons, $<0.5 $ events,) can be distinguished from the signal 
with sufficiently high efficiency or are negligible above the 
2.6 MeV analysis threshold (table I in 
Ref. \cite{kloctober}).
We also consider negligible the background from geological neutrinos 
 above the 
2.6 MeV analysis threshold.

\subsection{The Solar Signal}

We  also  need  the expected signals in the 
different solar neutrino experiments.
These  are obtained by convoluting solar neutrino 
fluxes, sun  and earth 
oscillation probabilities, neutrino cross sections 
and detector energy response functions. We closely follow 
the same methods already well explained in previous works
\cite{Aliani:2001zi,Aliani:2001ba,Aliani:2002ma,Aliani:2002rv}, 
we will mention here only a few aspects 
of this computation.
We determine the neutrino oscillation probabilities 
using the standard
methods found in literature~\cite{torrente}, as explained 
in detail in~\cite{Aliani:2001zi} and in~\cite{Aliani:2002ma}. 
We use a thoroughly numerical method to calculate the 
neutrino evolution equations in the presence of matter for all 
the parameter space.
For the solar neutrino case 
the calculation is split in three steps, corresponding to 
the neutrino propagation inside the Sun, in the vacuum 
(where the propagation is computed analytically) and in the Earth.
We average over the neutrino production point inside the Sun 
and we take the electron number density $n_e$ in the Sun by the BPB2001 model~\cite{bpb2001}.
The averaging over the annual  variation of the orbit
is also exactly performed.
To take the Earth matter effects into account, we adopt a
 spherical model of the Earth  density and chemical composition~\cite{earthprofile}.
The joining of the neutrino propagation in the three different 
regions is performed exactly using an evolution operator
formalism~\cite{torrente}.
The final survival probabilities are obtained from the 
corresponding (non-pure) density matrices built from 
the evolution operators in  each of these three regions.

In this analysis in addition to night probabilities we will 
need the  partial night probabilities corresponding 
to the 6 zenith angle bin data presented by SK \cite{Smy:2002fs}.
They are obtained using appropriate weights 
 which depend on the neutrino  impact parameter and the 
 sagitta distance from neutrino trajectory  to the Earth's center,
 for each detector's geographical location.

\section{Analysis and Results}
\label{analysis}

In order to study power of the  Kamland results for 
resolving the neutrino oscillation parameter space,
we have developed  two kind of analysis. 
In the first case we  deal with the Kamland 
measured global signal. 
In the second case we  include the full 
Kamland spectrum information. 
We perform a complete $\chi^2$ 
statistical analysis before and after
 including in addition the 
up-date solar evidence
obtaining 
the regions in the parameter space which are favored.  

\subsection{Analysis of the Global rate}

The total $\chi^2$ value is given by the sum of 
two distinct contributions, that is the one coming from all 
the solar neutrino data and the contribution of the Kamland 
experiment:
$$  \chi^2= \chi^2_{\odot} + \chi^2_{glob,KL},$$
with
\begin{eqnarray}
  \chi^2_{glob,KL}&=& \left 
(\frac{R^{exp}-R^{teo}(\Delta m^2,\theta)}{\sigma_{stat+sys}}\right )^2.
\end{eqnarray}
The ``experimental'' signal ratio is $R^{exp}=0.611\pm 0.085\pm 0.041$
\cite{kloctober}, where the first error is statistical and the 
second one  systematic.

The solar neutrino contribution can be written in 
the following way:
\begin{eqnarray}
  \chi^2_{\odot }&=& \chi^2_{\rm glob}+\chi^2_{\rm SK}+\chi^2_{\rm SNO}.
\end{eqnarray}
The function $\chi^2_{\rm glob}$ 
correspond to the total event rates measured at the  
Homestake  experiment~\cite{Homestake} and at the gallium 
experiments SAGE~\cite{sage,sage1999}, 
GNO~\cite{gno2000} and GALLEX~\cite{gallex}. 
We follow closely the definition used in 
previous works (see Ref.\cite{Aliani:2002ma}
 for definitions and 
 Table~(1) in Ref.\cite{Aliani:2002ma} ` 
for an explicit list of results 
and other  references).

The contribution to the $\chi^2$ from the SuperKamiokande 
data ($\chi^2_{\rm SK}$) has been obtained by using  
  double-binned data in energy 
 and zenith angle (see table 2 in Ref.\cite{Smy:2002fs} and
 also Ref.\cite{Fukuda:2002pe}):
8 energy bins of variable width and 7 zenith angle bins which
 include the day bin and 6 night ones. The definition is 
given by: 
\begin{eqnarray}
\label{chi2}
  \chi^2_{\rm SK}
&=& ({\vR^{\rm th} -\vR^{\rm exp}})^t 
\left (\sigma^{2}_{\rm unc} + \sigma^{2}_{\rm cor}\right )^{-1}
 ({ \vR^{\rm th}-\vR^{\rm exp}}).
\end{eqnarray}
The theoretical and experimental $\vR$ quantities  
are this time matrices of dimension 8$\times$7. 
The covariance quantity $\sigma$ is a 4-rank 
tensor constructed  in terms of
statistic errors, energy and zenith angle 
bin-correlated and uncorrelated uncertainties.
The data and errors for individual energy 
 bins for SK spectrum has been obtained from Ref.~\cite{Smy:2002fs}. 

The contribution of SNO to the $\chi^2$ is given by
\begin{eqnarray}\label{chi2b}
  \chi^2_{\rm spec-SNO}
&=&\sum_{d,n} ({\vR^{\rm th} -\vR^{\rm exp}})^t 
\left (\sigma^{2}_{\rm stat} + \sigma^{2}_{\rm syst}\right )^{-1}
 ({\vR^{\rm th}-\vR^{\rm exp}}),
\label{chiall}
\end{eqnarray}
where the day and night $\vR$ vectors of dimension 17 
are made up by the
values of the total (NC+CC+ES) SNO signal for the 
different bins of the spectrum.  
The statistical contribution to the covariance 
matrix, $\sigma_{\rm stat}$,
is obtained directly from the SNO data. The part of the matrix 
related to the systematical errors  has been computed  
by us studying the influence on the response function 
of the different sources of 
correlated and uncorrelated errors reported by SNO 
collaboration (see table II of
 Ref.~\cite{Ahmad:2002jz}), we assume 
full correlation or full anticorrelation according to 
each source.

To test a particular oscillation hypothesis against the 
parameters of the best fit 
and obtain allowed regions in parameter space we perform a 
 minimization of the two dimensional function
 $\chi^2(\Delta m^2,\tan^2\theta)$. 
A given point in the oscillation parameter space is allowed if 
 the globally subtracted quantity 
fulfills the condition 
 $\Delta \chi^2=\chi^2 (\Delta m^2, \theta)-\chi_{\rm min}^2<\chi^2_n(CL)$.
Where $\chi^2_{n=2}(90\%,95\%,...)$ are the quantiles for
 two degrees of freedom.

\begin{figure}
\centering
\begin{tabular}{lcr}
\psfig{file=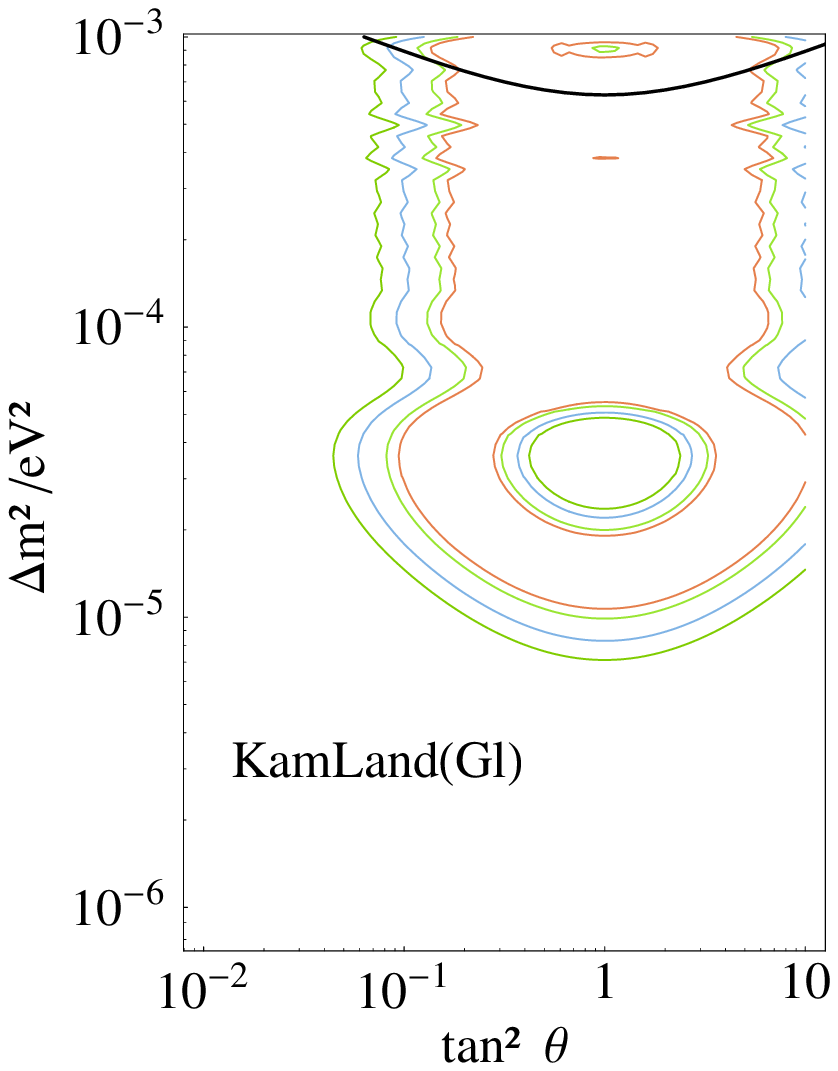,width=6.5cm}
&\hspace{1cm} &
\psfig{file=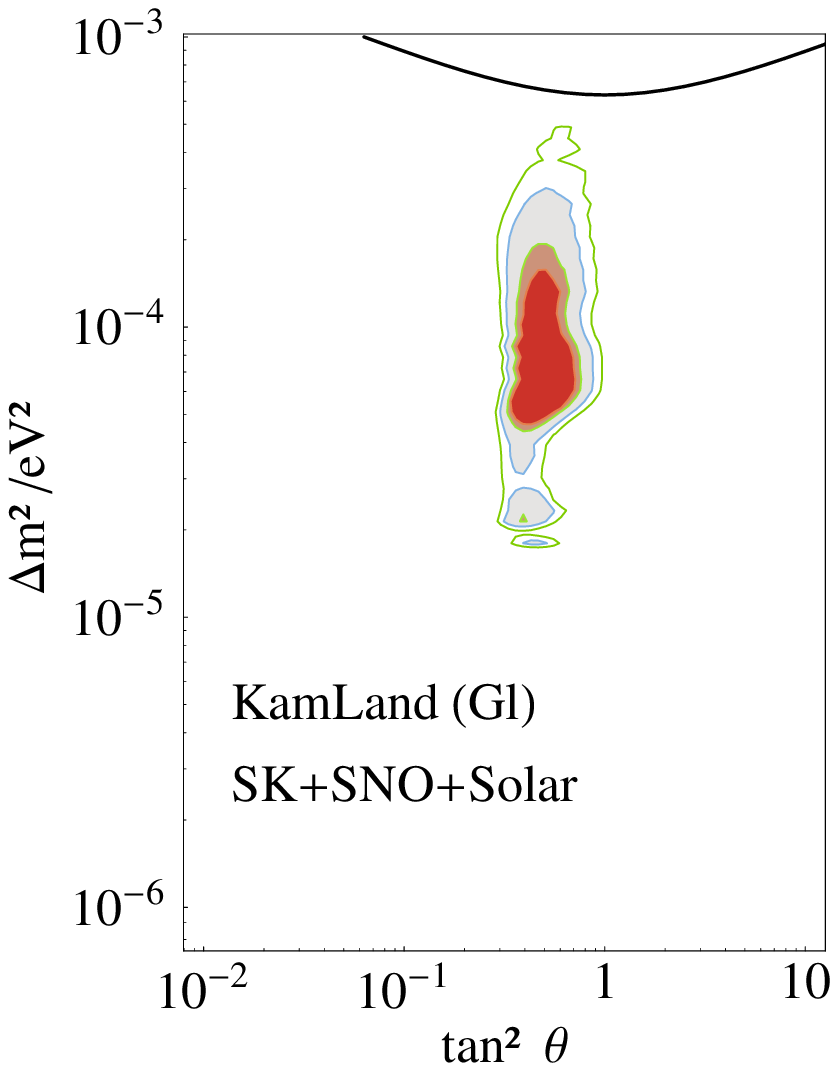,width=6.5cm}
\end{tabular}
\caption{
\small Exclusion plots including Kamland global rates.
The colored areas are allowed   at 
90, 95, 99 and 99.7\% CL relative to the absolute minimum.
The region above the upper thick line is excluded by the 
previous CHOOZ experiment \protect\cite{chooznew}.
}
\label{f1}
\end{figure}

In Figs.\ref{f1} we graphically show the 
results of this analysis. 
The favored region by the Kamland global rate alone is 
presented in fig.\ref{f1}(Left).
Fig.\ref{f1}(right) includes the full Kamland and Solar 
contributions.
Kamland global rate alone restricts the value of  
$\tan^2\theta$ to be in the range 
$\tan^2\theta > 4\times 10^{-1}-1$ 
(symmetric with respect
the value $\theta = \pi/4 $)
except for some small region corresponding to
 $\tan^2\theta \sim 1$ and 
$\Delta m^2 \sim 2-3\times 10^{-5}$ eV$^2$.
The overall effect of the measured observed ratio 
$R\sim 0.6\%$ is that the LMA region remains the only 
one which is still favored. 
Only if Kamland should had  
measured a ratio $\sim 0.1-0.3$,  more than three sigma 
away from the actual measurement, other regions as the 
LOW solar solution would have some statistical chance to 
survive.
In  Table~(I.a)
 we present the best fit parameters or local minima 
 obtained from the minimization of the $\chi^2$ function.
Also shown are the values of $\chi^2_{\rm min}$.
According to these results, regardless of the individual 
Kamland results,  a maximal mixing solution is excluded 
when solar data is included in the analysis.

\subsection{Analysis including the full Kamland signal }

Here we use the  binned Kamland signal
(See table 2 extracted from Fig.5
 in  Ref.\cite{kloctober})
combined with
the evidence of the up-date 
 solar experiments (CL,GA,SK and SNO).
The total $\chi^2$ value is given now by the 
sum of two distinct contributions, that is the one coming from 
all the solar neutrino data and the contribution of the 
Kamland experiment which includes both, shape and total signal information:
\begin{eqnarray}
  \chi^2&=& \chi^2_{\odot} + \chi^2_{spec,KL} + \chi^2_{glob,KL}.
\end{eqnarray}

The contributions of the solar neutrino experiments
$\chi^2_{\odot}$ and global Kamland signal 
 are described in detail in the previous section.
The contribution of the Kamland spectrum is now as follows:
\begin{eqnarray}
  \chi^2_{\rm spec,KL}&=& ({\alpha \vR^{\rm th}-\vR^{\rm exp}})^t 
\left (\sigma_{unc}^2+\sigma_{corr}^2 \right)^{-1} ({\alpha \vR^{\rm th}-\vR^{\rm exp}})
\label{chiklspec}
\end{eqnarray}

The total error matrix $\sigma$ is computed as a sum of assumed 
systematic deviations, $\sigma_{sys}/S\sim 6.5\%$,
 mainly coming from flux uncertainty 
($3\%$), energy calibration and threshold
(see table II of
 Ref.~\cite{kloctober} for a total systematic error
$\sim 6.4\%$), see also Ref.\cite{klstony,Murayama:2000iq,kamlandmonaco}) 
and statistical errors. The parameter $\alpha$ is a free normalization parameter.
The   effect of systematic sources on 
individual bin deviations  has been computed  
by us studying the influence on the response function, 
furtherly 
we have assumed full correlation among bins.

Note that, as an alternative to the previous formula $\ref{chiklspec}$ 
based on ratios and using and standard $\chi^2$ expression, we 
could have directly used the absolute number of events
in each bin and applied
 poissonian statistics formulas.
The Kamland collaboration in its published analysis uses a 
mixed strategy and
 divides the $\chi^2$ expression in two parts: it adds a 
standard  $\chi^2$ contribution corresponding to the 
global rate to a likelihood  contribution which models the 
statistical significance of the spectrum shape.
In this work, where we are interested not only on the Kamland 
data itself but in combining it with the solar data, we prefer 
to continue using the familiar expression \ref{chiklspec} for 
various reasons: first, for simplicity, clarity and comparability 
with previous and future results, second,  in the major part of 
the bins, certainly in the most significant ones, the signal is high 
enough, and more importantly the errors are 
small enough, to use the Gaussian approximation. For the large 
energy bins the signal is strictly zero: its contribution is 
anyway effectively zero in both approaches. 
Finally, not less importantly, the possibility of introducing in our 
approach in a straightforward way the effect of correlated
systematic deviations among bins.

The ${\vR}$ are length 13 vectors containing the binned 
spectrum (0.425 MeV bins ranging from 2.6 to 8.125 MeV) 
normalized to the non-oscillation 
 expectations. Theoretical vectors are a function of the 
oscillation parameters:
$\vR^{\rm th}=\vR^{\rm th}(\Delta m^2,\theta)$. 
 The experimental vectors
$\vR^{\rm exp}$ contain the Kamland binned signal.
We generate acceptance contours
 (at 90,95 and 99 \% CL)  in the 
$(\Delta m^2,\tan^2 \theta)$ plane in a similar manner 
as explained in the previous section including now a minimization respec the 
parameter $\alpha$ for any of the other oscillation parameters.
 For the sake of 
comparison we have also obtained exclusion regions derived 
from the consideration of the Kamland contribution 
alone ($\chi^2_{KL}=\chi^2_{spec}+\chi^2_{gl}$).

\begin{figure}
\centering
\begin{tabular}{rcl}
\psfig{file=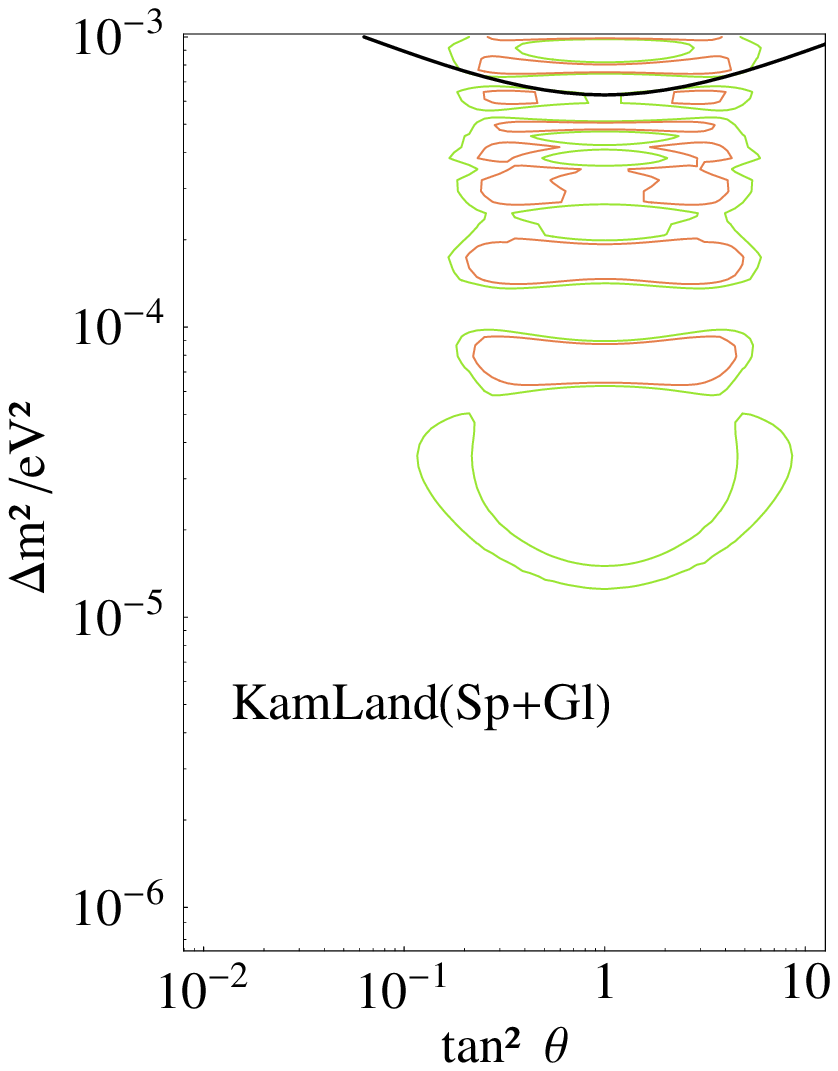,width=6.5cm}
&\hspace{1cm} &\psfig{file=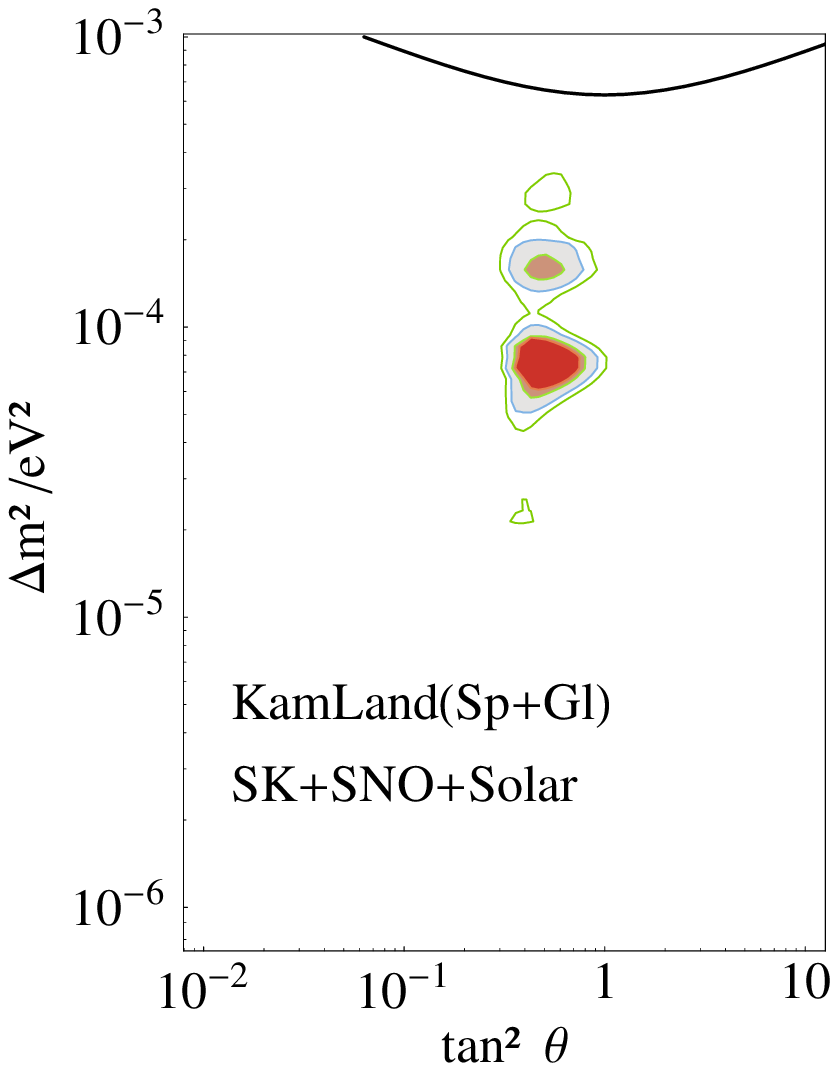,width=6.5cm}
\end{tabular}
\caption{\small 
Allowed areas in the two neutrino parameter space
after 150 days of data taking in Kamland.
The colored lines separate allowed regions at 
90, 95, 99 and 99.7\% CL relative to the absolute minimum.
(Left) Results with the Kamland signal alone (shape and total signal information, 
only 90,95\% Cl lines 
are shown for clarity).
(Right) Kamland spectrum plus solar (CL,GA,SK,SNO) evidence.
}
\label{f2}
\end{figure}

In Figs.(\ref{f2})  we graphically show the results of this analysis.
The first case, study of the Kamland data alone ($\chi^2_{KL}$) is represented by the 
 Fig.(\ref{f2})( left). The allowed regions in 
parameter space corresponding to each particular point is formed by a number of 
regions symmetric with respect the line $\tan\theta=1$.
The general effect of the inclusion of the solar 
evidence, showed in Fig.\ref{f2} (right), 
in the $\chi^2$ is the breaking of the symmetry in 
$\tan^2 \theta$, as expected, and the strong contraction of the 
allowed area to well defined regions.
The best fit parameters 
 obtained from the minimization of the $\chi^2$ function are presented 
in   Table~(I.a) where two well separated solutions appear LMAI,LMAII.
In both cases the mixing angle is around $\tan^2\theta\sim 0.45-0.48$. 

Again, the introduction of the solar data strongly diminishes the favored value for the 
mixing angle. The final value is  more near to those values favored by the solar data 
alone than to the Kamland ones. This effect could be simply due to the present low 
Kamland statistics or, more worrying, to some statistical artifact derived from the complexity 
of the analysis  and of the heterogeneity of binned data involved. 

\begin{figure}
\centering
\begin{tabular}{c}
\psfig{file=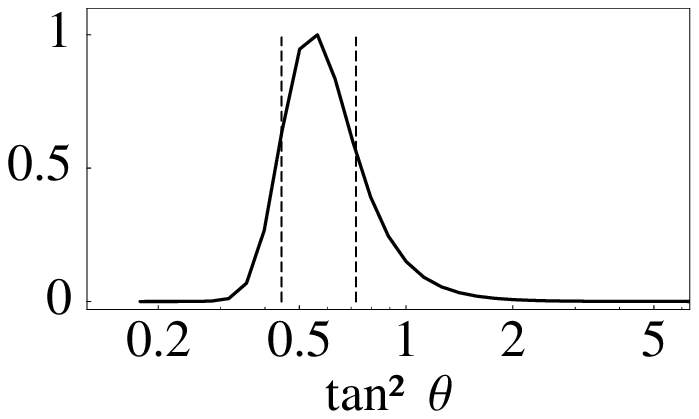}\\
\psfig{file=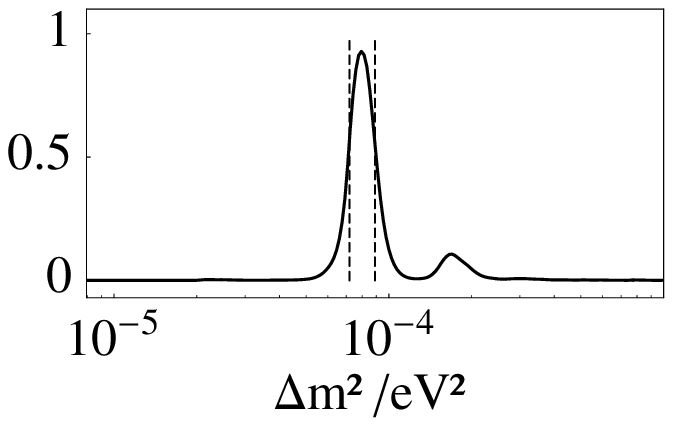}
\end{tabular}
\caption{\small  
Marginalized likelihood distributions for each of the
 oscillation parameters 
$\Delta m^2$ (right), $\tan^2 \theta$ (left)
corresponding to the Kamland Spectrum plus  solar evidence
 (Fig.\protect\ref{f2}(Right)).
The curves are in arbitrary units with normalization to the 
 maximum height.
Values for the peak position are obtained by fitting 
two-sided Gaussian distributions (not showed in the plot).
Dashed lines delimit $\pm 1\sigma$ error regions around the maximum. See Table I.a for values of the position and widths of 
the peaks.}
\label{f3}
\end{figure}

In this respect, we perform additionally 
a second kind of analysis in order to obtain concrete values for the 
individual oscillation parameters and estimates for their uncertainties. 
We study the marginalized parameter constraints where the  $\chi^2$ quantity is converted 
into likelihood using the expression ${\cal L}
=e^{-(\chi^2-\chi_{min}^2)/2}$. 
This normalized marginal likelihood, obtained from the integration of ${\cal L}$ for each 
of the variables, is plotted in Figs.~(\ref{f3}) for each of the oscillation 
parameters $\Delta m^2$ and $\tan^2\theta$. For $\tan^2\theta$ we observe that the 
likelihood function is strongly peaked  in a 
 region $\tan^2\theta\sim 0.4-0.6$.
The situation for $\Delta m^2$ is similar except for the existence of additional, somehow narrower, 
secondary peaks. Concrete values for the parameters are extracted by fitting  one- or two-sided 
Gaussian distributions to any of the peaks (fits not showed in the plots). In both cases, for  angle 
and the mass difference distributions the goodness of fit of the Gaussian fit to each individual peak 
is excellent (g.o.f $>99.8\%$) thus justifying the consistency of the procedure. The values for the 
parameters obtained in this way appear in Table~(I.b). They are fully consistent and very similar to 
the values obtained from simple $\chi^2$ minimization. In particular, the maximal mixing 
solution is again excluded at the $\sim 3\sigma$ level. 

Although both are mutually compatible, the slight difference of the value obtained 
for the mixing angle is well explained by the shape of the allowed regions in 
Fig~\ref{f2} (right): the right elongation of these makes the value of the integral which defines 
the marginal distribution for $\tan^2\theta$ to be shifted. Additional variability can be easily 
introduced if would have used different prior information or different parameter 
definition (i.e. $\sin 2\theta$ instead of $\tan^2\theta$).  This an example of how the details of 
statistical analysis can significantly modify the values of the physical parameters extracted from 
data and how important is that different methods are explored, specially in the present context of 
neutrino physics where heterogeneous data coming from very different sources is jointly used.

\section{ Summary and  Conclusions}\label{sec:conclusions}
\label{conclusions}

We have analyzed the present experimental situation of our 
knowledge of the neutrino mixing parameters in the region 
of the 
parameter space that is relevant for solar neutrinos and 
we have studied in detail how this knowledge has improved 
with the recent results presented by the 
reactor experiment Kamland.
We show, how, in general,
 the regions selected by Kamland alone,  all symmetric with
 respect to $\tan^2 \theta =1$, have a large 
spreadth in the mixing angle.
The experiment has, however, a much higher sensitivity 
to the mass difference parameter.

Kamland global rate alone restricts the value of  
$\tan^2\theta$ to be in the range 
$\tan^2\theta > 4\times 10^{-1}-1$ 
(symmetric with respect
the value $\theta = \pi/4 $)
except for some small region corresponding to
 $\tan^2\theta \sim 1$ and 
$\Delta m^2 \sim 2-3\times 10^{-5}$ eV$^2$.
The overall effect of the measured observed ratio 
$R\sim 0.6\%$ is that the LMA region remains the only 
one which is still favored. 
Only if Kamland should had  
measured a ratio $\sim 0.1-0.3$,  more than three sigma 
away from the actual measurement, other regions as the 
LOW solar solution would have some statistical chance to 
survive.

From the analysis of the  Kamland evidence alone,
 the allowed regions in parameter space corresponding to each 
particular point is formed by a number regions symmetric with respect the 
line $\tan\theta=1$. The general effect of the inclusion of the solar 
evidence in the analysis  is the breaking of the symmetry in 
$\tan^2 \theta$,  the strong reduction of the area of the allowed regions and, 
more importantly, the shifting of the value of the best fit mixing angle 
from being compatible with maximal mixing to more than 
$3\sigma$ incompatibility.

In addition to parameter extraction from  $\chi^2$ minimization, we 
obtain concrete values for the individual oscillation parameters and 
estimates for their uncertainties from the marginalized likelihood distributions.
For $\tan^2\theta$ we observe that the likelihood function is concentrated in a  region 
slightly higher that that one obtained from minimization.
According to this analysis, maximal mixing is again excluded at the $3\sigma$ level.
The marginal distribution for $\Delta m^2$ shows clearly the 
existence of an additional, somehow narrower and much less significant,
 secondary peaks. 
Kamland, after only 150 days of data taking, together with 
the rest of solar experiments is already 
able to resolve the neutrino mass difference with
 a high precision of 
$\delta  \Delta m^2_{kl+solar}< \pm 0.8$  to be 
compared 
with the solar only case where the 
precision obtained was
$\delta  \Delta m^2_{solar}< \pm 1.5$.
The situation for the other oscillation 
parameter is worse, from the present data we obtain
 $\delta \tan^2\theta_{kl+solar} < \pm 0.15$. 
This precision
is comparable, in fact even worse, than the 
precision attainable from the solar data alone 
 $\delta \tan^2\theta_{solar} < \pm 0.10$ \cite{Aliani:2002ma}.
The perspectives of Kamland, after 1-3 years of data taking, 
together or not with 
solar evidence, for a much better determination 
of the mixing angle are not very optimistic 
(see the results in Ref.\cite{kltorrente}). 
The  ability of  experiments as BOREXINO to 
improve the determination of the mixing angle with 
a smaller error in the near future remains an open 
question (see Ref.\cite{torrentebor}).

In summary, in the framework of two active neutrino oscillations 
we obtain, combining Kamland and Solar data and assuming CPT invariance, 
i.e. the same mass matrix for neutrino and antineutrinos, 
the following values for  neutrino mixing parameters
$ \Delta m^2_{kl}= 7.1\times 10^{-5} \eV^2,
\tan^2\theta_{kl}=  0.47$ from $\chi^2$ minimization (LMAI solution).

\vspace{0.3cm}
\subsection*{Acknowledgments}

We  acknowledge the  financial  support of 
 the Italian MURST and  the  Spanish CYCIT  funding 
agencies. 
The numerical calculations have 
been performed in the computer farm of 
 the Milano University theoretical group.

\newpage

\begin{table}[p]
 \begin{center}

  \scalebox{0.9}{
    \begin{tabular}{llll}
 & $\Delta m^2 (\eV^2) $& $\tan^2\theta$&  $\chi_{min}^2$ 
\\[0.15cm]
\hline
Table I.a (Minimization $\chi^2$): &   &   & \\[0.17cm]
\hspace{0.3cm} KL (Sp)&  $7.7\times 10^{-5}$ &   $0.98$    & 1.93  \\[0.15cm]
\hspace{0.3cm} KL (Gl)+Solar& $5.8\times 10^{-5}$ & $0.47$ & 45.8    \\[0.15cm]
\hspace{0.3cm} KL (Sp+Gl)+Solar, LMAI&  $7.1\times 10^{-5}$ & $0.47$ & 49.1  \\[0.15cm]
\hspace{0.3cm} KL (Sp+Gl)+Solar, LMAII&  $1.5\times 10^{-4}$ & $0.48$ & 49.8  \\[0.25cm]
 Table I.b ( From Fit, Fig.\protect\ref{f3}, $\pm 1\sigma$): &   &   & \\[0.17cm]
\hspace{0.3cm} KL (SP+Gl)+Solar, LMAI & $8.0^{+0.9}_{-0.8}\times 10^{-5} $ &  &  \\[0.15cm]
\hspace{0.3cm} KL (SP+Gl)+Solar, LMAII & $1.7^{+0.2}_{-0.2}\times 10^{-4} $ &  & \\[0.15cm]
\hspace{0.3cm} KL (SP+Gl)+Solar  &  & $0.55^{+0.16}_{-0.12}$   & \\[0.15cm]
\hline
      \end{tabular}
}

 \caption{\small
Mixing parameters: 
 From $\chi^2$ minimization (Tables I.a) and 
from fit to the peak of marginal likelihood distributions 
(Table I.b).
}
  \label{table1}
 \end{center}
\end{table}

\begin{table}
\centering
\begin{tabular}{lcl}
\hline \hline
Bin (MeV) &   $R=S_{exp}/S_{MC}$& $\pm 1\sigma_{stat}$ \\[0.15cm]
\hline
2.600-3.025 &  $0.435  $ &    $\pm 0.160 $            \\
3.025-3.450 &  $0.689  $ &    $\pm 0.215$            \\
3.450-3.875 &  $0.666  $ &    $\pm 0.225$            \\
3.875-4.300 &  $0.719  $ &    $\pm 0.250$            \\
4.300-4.725 &  $0.885  $ &    $\pm 0.310$            \\
4.725-5.150 &  $0.550  $ &    $\pm 0.305$            \\
5.150-5.575 &  $1.000  $ &    $\pm 0.460$            \\
5.575-6.000 &  $0.598  $ &    $\pm 0.500$            \\
6.000-6.425 &  $0.000  $ &    $\pm 0.365$            \\
6.425-6.850 &  $0.000  $ &    $\pm 0.630$            \\
6.850-7.275 &  $0.000  $ &    $\pm 2.500$            \\
7.275-7.700 &  $0.000  $ &    $\pm 2.500$            \\
7.700-8.125 &  $0.000  $ &    $\pm 2.500$            \\[0.15cm]
 \hline\\
\end{tabular}
\caption{Summary of Kamland signal ratios as used in the
present work. The table has been obtained from the 
information contained in
Fig.5 in Ref.\protect\cite{kloctober}.}
\label{t2}
\end{table}

\clearpage

\vspace{1cm}

\end{document}